\documentclass[9pt,letterpaper,aps,pra,notitlepage,twocolumn,reprint,shortbibliography]{revtex4-1}
\usepackage[utf8]{inputenc}
\usepackage{amsmath,amssymb}
\usepackage{graphicx}
\usepackage[margin=2cm]{geometry}
\usepackage{bm}

\newcommand{\eq}[1]{\begin{align}#1\end{align}}

\newcommand{\Tor}{\text{Tor}}

\usepackage{newfloat}

\usepackage{algorithmic}
\usepackage{algorithm}
\usepackage{color}
\usepackage{bbm}
\usepackage{comment}
\DeclareMathOperator{\one}{\mathbbm{1}}

\newcommand{\beq}{\begin{equation}}
\newcommand{\eeq}{\end{equation}}

\definecolor{JM}{RGB}{4,116,149}

\definecolor{nico}{RGB}{51, 87, 255}

\begin{document}
	\title{Exact simulation of Gaussian Boson Sampling in polynomial space and exponential time}
	\author{Nicol\'as Quesada}
	\email{nicolas@xanadu.ai}
	\affiliation{Xanadu, 777 Bay Street, Toronto, Canada}
	\author{Juan Miguel Arrazola}
	\affiliation{Xanadu, 777 Bay Street, Toronto, Canada}
	
\begin{abstract}
We introduce an exact classical algorithm for simulating Gaussian Boson Sampling (GBS). The complexity of the algorithm is exponential in the number of photons detected, which is itself a random variable. For a fixed number of modes, the complexity is in fact equivalent to that of calculating output probabilities, up to constant prefactors. The simulation algorithm can be extended to other models such as GBS with threshold detectors, GBS with displacements, and sampling linear combinations of Gaussian states. In the specific case of encoding non-negative matrices into a GBS device, our method leads to an approximate sampling algorithm with polynomial runtime. We implement the algorithm, making the code publicly available as part of Xanadu's The Walrus library, and benchmark its performance on GBS with random Haar interferometers and with encoded Erd\H{o}s-Renyi graphs. 

\end{abstract}

\maketitle
\section{Introduction}
Boson Sampling is a model of photonic quantum computing that was introduced to argue that that non-universal photonic quantum computers cannot be efficiently simulated classically \cite{aaronson2011computational}. Since then, significant work has been done to pursue its practical implementation \cite{tillmann2013experimental,spring2013boson,broome2013photonic,bentivegna2015experimental} and to design variants that are more amenable to existing experimental techniques \cite{barkhofen2017driven,lund2014boson,lund2017exact,chakhmakhchyan2017boson}. Recently, Gaussian Boson Sampling (GBS) has emerged as a new paradigm that addresses some of the major challenges in scaling Boson Sampling devices \cite{hamilton2017gaussian,kruse2018detailed}. Instead of employing indistinguishable single photons as inputs, GBS prepares a multi-mode Gaussian state that is subsequently measured using photon detectors \cite{rosenberg2005noise}. The Gaussian state is typically obtained by sending squeezed light through a linear-optical interferometer, while more general versions employ displacements together with squeezing operations. GBS has raised additional interest due to the discovery of applications to quantum chemistry \cite{huh2015boson}, optimization \cite{arrazola2018using, arrazola2018quantum, banchi2019molecular}, graph similarity \cite{ schuld2019quantum}, and point processes \cite{jahangiri2019point}. Initial experimental implementations have also been recently reported \cite{clements2018approximating, paesani2019generation, zhong2019experimental}.

The complexity arguments underlying the hardness of classically simulating GBS are only valid asymptotically. In practice, it is necessary to perform a comparison to state-of-the-art simulation algorithms to understand the actual advantages of employing a quantum computer. Considerable progress has been made in developing classical simulation algorithms for the original Boson Sampling model. Ref. \cite{neville2017classical} reported the first simulation method by describing an approximate Markov chain Monte Carlo algorithm for Boson Sampling. These results were improved in Ref. \cite{clifford2018classical}, where an exact sampling algorithm was developed with the same asymptotic complexity but better constant prefactors. An algorithm for other models of Boson Sampling has also been recently reported \cite{renema2019simulability}. As stated in Ref. \cite{moylett2019classically}, extending these techniques to GBS has been challenging because these algorithms rely on specific properties of Boson Sampling that are not present in GBS. Despite these challenges, an exact simulation algorithm has been reported and implemented for the specific case of GBS with threshold detectors \cite{quesada2018gaussian,gupt2018classical}. This algorithm suffers from the critical drawback that its memory requirement also scales exponentially, limiting the scope of problems that can be simulated.

In this work, we introduce an exact simulation algorithm for GBS with efficient space complexity and applicable to all versions of GBS. The worst-case time complexity of the algorithm is linear in the number of modes and exponential in the number of detected photons. The core strategy is to employ the chain rule of probability to sequentially sample the number of photons in each mode conditioned on the results from previous modes. Conditional sampling is particularly well suited to Gaussian states because it is straightforward to describe the marginalized density matrices of any number of subsystems using covariance matrices and a vector of means.
The method presented here can also be applied to several other models,  including GBS with threshold detectors, GBS with displacements, and sampling linear combinations of Gaussian states. We also show that in the special case of encoding non-negative matrices into a GBS device, the algorithm results in an approximate sampling procedure with polynomial runtime. Finally, we implement the algorithm and benchmark its performance on random interferometers drawn from the Haar measure. We also benchmark the approximate algorithm for random Erd\H{o}s-Renyi graphs. The code used to generate these results is freely available as part of Xanadu's The Walrus library \cite{code}.

\section{Classical GBS Algorithm}
In this section we describe and study the algorithm for GBS simulation. For completeness, we begin with a brief description of GBS. We continue by outlining the algorithm in detail, performing a theoretical analysis of its complexity, and discussing its scope of application. 

\subsection{Gaussian Boson Sampling}
The quantum state $\varrho$ of a system of $m$ bosonic modes can be uniquely specified by its Wigner function $W(\bm{q},\bm{p})$ \cite{weedbrook2012gaussian,serafini2017quantum}, where $\bm{q} \in \mathbb{R}^m$ are the canonical positions and $\bm{p} \in \mathbb{R}^m$ are the canonical momenta. Gaussian states are defined simply as the set of states with Gaussian Wigner functions. They can be uniquely described by a covariance matrix $\bm{V}$ and a vector of means $\bm{\bar{q}},\bm{\bar{p}}$. It is convenient to write the covariance matrix in terms of the complex amplitudes $\bm{\alpha} = \tfrac{1}{\sqrt{2 \hbar }} (\bm{q}+i \bm{p}) \in \mathbb{C}^m$. The variables $\bm{\alpha}$ are complex-normal distributed with mean $\bm{\bar{\alpha}} = \tfrac{1}{\sqrt{2 \hbar}} (\bm{\bar{q}}+i \bm{\bar{p}}) \in \mathbb{C}^m$ and covariance matrix ${\bm{\Sigma}}$  \cite{picinbono1996second}.

GBS is a model of photonic quantum computing where a Gaussian state is measured in the Fock basis. A general pure Gaussian state can be prepared by using single-mode squeezing and displacement operations in combination with linear-optical interferometry \cite{serafini2017quantum,clements2016optimal,de2018simple,reck1994experimental}. It was shown in Ref.~\cite{hamilton2017gaussian} that when the modes of a Gaussian state with zero mean ($\bm{\bar{\alpha}} = 0$) are measured, the probability of obtaining a pattern of photons $S = (s_1,\ldots,s_m)$, where $s_i$ is the number of photons in mode $i$, is given by
\begin{align}\label{eq:GBS}
p(S) = \frac{1}{\sqrt{\text{det}(\bm{Q})} } \frac{\text{Haf}(\bm{A}_S)}{s_1!\ldots s_m!}, 
\end{align}
where 
\begin{align}
\bm{Q}&:=\bm{\Sigma} +\mathbbm{1}/2,\\
\bm{A} &:= \bm{X} \left(\mathbbm{1} - \bm{Q}^{-1}\right),\label{Eq:A_matrix}\\
\bm{X} &:=  \left[\begin{smallmatrix}
	0 &  \one \\
	\one & 0  
\end{smallmatrix} \right],
\end{align}
and $\bm{A}_S$ is the matrix obtained by repeating rows and columns $i$ and $i+m$ of the $2m\times 2m$ symmetric matrix $\bm{A}$. We refer to $\bm{A}$ as the \emph{kernel} matrix. If $s_i=0$, rows and columns $i$ and $i+m$ are deleted from $\bm{A}$; if $s_i>0$, the rows and columns are repeated $s_i$ times. For example, consider $m=3$ modes and let
	\begin{align}
	\bm{A}=\left[
	\begin{array}{ccc|ccc}
	a_{1,1} & a_{1,2} & a_{1,3} & a_{1,4} & a_{1,5} & a_{1,6} \\
	a_{2,1} & a_{2,2} & a_{2,3} & a_{2,4} & a_{2,5} & a_{2,6} \\
	a_{3,1} & a_{3,2} & a_{3,3} & a_{3,4} & a_{3,5} & a_{3,6} \\
	\hline
	a_{4,1} & a_{4,2} & a_{4,3} & a_{4,4} & a_{4,5} & a_{4,6} \\
	a_{5,1} & a_{5,2} & a_{5,3} & a_{5,4} & a_{5,5} & a_{5,6} \\
	a_{6,1} & a_{6,2} & a_{6,3} & a_{6,4} & a_{6,5} & a_{6,6} \\
	\end{array}
	\right].
	\end{align}
	For the photon pattern $S = (3,0,1)$, one has
	\begin{align}
	\bm{A}_S = \left[
	\begin{array}{cccc|cccc}
	a_{1,1} & a_{1,1} & a_{1,1} & a_{1,3} & a_{1,4} & a_{1,4} & a_{1,4} & a_{1,6}
	\\
	a_{1,1} & a_{1,1} & a_{1,1} & a_{1,3} & a_{1,4} & a_{1,4} & a_{1,4} & a_{1,6}
	\\
	a_{1,1} & a_{1,1} & a_{1,1} & a_{1,3} & a_{1,4} & a_{1,4} & a_{1,4} & a_{1,6}
	\\
	a_{3,1} & a_{3,1} & a_{3,1} & a_{3,3} &	a_{3,4} & a_{3,4} & a_{3,4} & a_{3,6}
	\\
	\hline 
	a_{4,1} & a_{4,1} & a_{4,1} & a_{4,3} &	a_{4,4} & a_{4,4} & a_{4,4} & a_{4,6}
	\\
	a_{4,1} & a_{4,1} & a_{4,1} & a_{4,3} &	a_{4,4} & a_{4,4} & a_{4,4} & a_{4,6}
	\\
	a_{4,1} & a_{4,1} & a_{4,1} & a_{4,3} &	a_{4,4} & a_{4,4} & a_{4,4} & a_{4,6}
	\\
	a_{6,1} & a_{6,1} & a_{6,1} & a_{6,3} &	a_{6,4} & a_{6,4} & a_{6,4} & a_{6,6}
	\\
	\end{array}
	\right].
	\end{align}
	Note that the first and fourth rows and columns of the original matrix are repeated three times, the second and fifth columns disappeared and the third and sixth column of $\bm{A}$ appear only once.
	Finally, the matrix function ${\rm Haf}(\cdot)$ is the 
hafnian \cite{caianiello1953quantum}, which for a $2m\times 2m$ matrix $\bm{A}$ is defined as
\begin{equation}
{\rm Haf}(\bm{A}) = \sum_{\mu\in {\rm PMP}} \prod_{(i,j)\in \mu} a_{i,j},
\end{equation}
where $a_{i, j}$ are the entries of $\bm{A}$ and $\rm PMP$ is the set of perfect matching permutations, namely 
the possible ways of partitioning the set $\{1,\dots,2m\}$ into disjoints subsets of size two. The hafnian is \#P-Hard to approximate for worst-case instances  \cite{barvinok2016combinatorics} and the runtime of the best known algorithms for computing hafnians scales exponentially with the dimension of the input matrix \cite{bjorklund2018faster}. The hardness of computing hafnians has been leveraged to show that, under the validity of specific technical conjectures, sampling from the output distribution of a GBS device cannot be performed efficiently using classical computers \cite{aaronson2011computational, hamilton2017gaussian}.

\subsection{Algorithm}
Ideally, a classical sampling algorithm for GBS will have the following properties: (i) sampling from the GBS distribution is \emph{exact}, (ii) space complexity is polynomial, and (iii) time complexity is proportional to the complexity of computing output probabilities. We now describe an algorithm satisfying all these properties. The main strategy is to apply the definition of conditional probability $p(s_k|s_{k-1},\cdots, s_{1})=p(s_k,s_{k-1},\cdots, s_{1})/p(s_{k-1},\cdots, s_{1})$ to sequentially sample each mode conditioned on outcomes from previous modes. When sampling the $k$-th mode, the probability $p(s_k,s_{k-1},\cdots, s_{1})$ can be calculated from the reduced state of the first $k$ modes, which can be efficiently computed for Gaussian states. This in turn involves calculating the hafnian of a matrix whose size depends on the number of photons detected so far. The probability $p(s_{k-1},\cdots, s_{1})$ is calculated from the previous step. 

Formally, let $\bm{V}^{(k)}$ denote the reduced covariance matrix of the first $k$ modes; this is simply the submatrix of $\bm{V}$ obtained by keeping rows and columns $1$ to $k$ and $m+1$ to $m+k$. From the reduced covariance matrix we can contruct $\bm{Q}^{(k)} = \bm{V}^{(k)}+\one/2$, $\bm{O}^{(k)} = \one - \left( \bm{Q}^{(k)}\right)^{-1}$ and $\bm{A}^{(k)} = \bm{X} \bm{O}^{(k)}$. The probability of observing a partial pattern $S^{(k)}=(s_1, s_2, \ldots, s_k)$ of photons in the first $k$ modes is given by
\beq\label{Eq:p(s)}
p(S^{(k)}) = \frac{1}{\sqrt{\text{det}(\bm{Q}^{(k)})} } \frac{\text{Haf}\left(\bm{A}^{(k)}_s\right)}{s_1!\ldots s_k!}.
\eeq

The algorithm is specified as follows:
\begin{enumerate}
\item Compute $\bm{A}^{(1)}$ and sample the number of photons $s_1$ in the first mode from the distribution $p(s_1) = \frac{1}{\sqrt{\text{det}(\bm{Q}^{(1)})} } \frac{\text{Haf}(\bm{A}^{(1)}_{s_1})}{s_1!}$.
This is done by computing each probability $p(s_1=0),p(s_1=1), \ldots, p(s_1=n_{\text{max}})$ up to a maximum photon number $n_{\text{max}}$ and sampling from the resulting distribution. The parameter $n_{\text{max}}$ must be chosen to ensure that the resulting distribution is sufficiently close to normalized, as discussed in Appendix \ref{Sec:Appendix2}. Let $s_1^*$ denote the output of this step.
\item Compute $\bm{A}^{(2)}$ and, as before, sample the number of photons $s_2$ in the second mode from the conditional distribution 
\beq
p(s_2|s_1^*) = \frac{p(s_1^*, s_2)}{p(s_1^*)},
\eeq
where $s_1^*$ is fixed from the result of the previous step and $p(s_1^*, s_2)$ is defined as in Eq.~\eqref{Eq:p(s)}.
\item Repeat this procedure for all modes. The conditional distribution of the photon number $s_k$ for the $k$-th mode is
\beq
p(s_k|s_1^*,\ldots,s_{k-1}^*)=\frac{p(s_1^*, s_2^*, \ldots,s_{k-1}^*, s_k)}{p(s_1^*,\ldots,s_{k-1}^*)}.
\eeq
The outputs $(s_1^*,\ldots,s_{k-1}^*)$ are fixed from the results of the previous steps, $p(s_1^*, s_2^*, \ldots,s_{k-1}^*, s_k)$ is given by Eq.~\eqref{Eq:p(s)}, and the probability $p(s_1^*,\ldots,s_{k-1}^*)$ that is necessary to sample at step $k$ has already been computed at step $k-1$.
\end{enumerate}

The correctness of the algorithm follows from the chain rule of probability
\beq
p(s_1, s_2, \ldots, s_m)=\prod_{k=1}^m p(s_k|s_1, s_2, \ldots, s_{k-1}),
\eeq
which implies that the algorithm performs exact sampling from the GBS distribution. 

\subsection{Complexity} \label{Sec: complexity}
The algorithm has $O(m^2)$ space complexity: the largest memory requirement arises from the need to store the $2m\times 2m$ matrix $\bm{A}$. Therefore, the algorithm is memory-efficient. To determine the time complexity, consider step $k$  where the goal is to sample photons from the $k$-th mode. Assume that $n$ photons have been sampled in the previous $k-1$ modes. To generate a sample, the algorithm computes the $n_{\text{max}}$ conditional probabilities of observing $0,1,\ldots,n_{\text{max}}$ photons in mode $k$. This in turn requires calculating the hafnian of matrices with dimensions $2n, 2(n+1), \ldots, 2(n+n_{\text{max}})$. State-of-the-art algorithms for evaluating hafnians \cite{bjorklund2018faster} have complexity $O(n^32^n)$, where $2n$ is the dimension of the matrix. Therefore, having detected $n$ photons, the time $t(n)$ required to sample one additional mode is
\begin{align}
t(n) &= O\left(\sum_{j=0}^{n_{\text{max}}}(n+j)^32^{(n+j)}\right)\nonumber\\
&=O\left(n^32^{n}\right).
\end{align} 
After each new step, the number of photons detected at that stage of the algorithm may either increase if more photons are observed, or stay the same if no more photons are detected. Let $\mu_n$ denote the number of steps for which exactly $n$ photons were detected at that stage of the algorithm. Note that $\mu_n\leq m$. To produce a sample of $N$ photons in total, the algorithm therefore requires time
\begin{align}
T(N, m) &= \sum_{n=0}^N\mu_n t(n)\nonumber\\
&= O\left(m N^32^{N}\right).
\end{align}
The largest matrix whose hafnian needs to be computed has dimension precisely $2(N+n_{\text{max}})$, so computing its hafnian takes time $(N+n_{\text{max}})^3 2^{N+n_{\text{max}}}=O(N^32^N)$. Thus, up to constant prefactors, the complexity of generating one sample from the algorithm is roughly $m$ times that of computing one output probability.

The complexity of the algorithm depends on how many photons have been sampled previously: two samples with the same number of photons can take different amounts of time to be generated. The two extreme cases occur when most of the photons are detected in the first few modes (longest time) and when they are detected in the last few modes (shortest time). When $N$ photons are measured in the first $N$ modes, the algorithm must compute $m-N$ hafnians of matrices with maximum dimension $2(N+n_{\text{max}})$. Conversely, if the $N$ photons are detected in the last $N$ modes, the algorithm calculates hafnians of the smallest possible dimension $2n_{\text{max}}$ in the first $m-N$ modes.

Although asymptotically the best algorithms for computing hafnians have complexity $O(N^3 2^N)$, the constant factors are sufficiently large that for all values of $N$ for which hafnians can be computed in practice, we note that it is in fact preferable to employ the algorithm of Ref.~\cite{bjorklund2012counting}, whose runtime scales as $O(N^52^N)$.

\subsection{Scope of application}

The sampling algorithm relies on an important property of Gaussian states: the marginal of a Gaussian state is another Gaussian state whose covariance matrix and vector of means can be efficiently computed from those of the larger state.
This is a special feature of Gaussian states; indeed, computing reduced states generally requires exponential time.
Using this observation, we generalize the sampling algorithm to several variants of GBS where marginals can be efficiently computed.

\subsubsection{Threshold detectors} The first case we consider is GBS with threshold detectors, where output probabilities are given by
\beq
p\left(S^{(k)}\right) = \frac{\Tor\left(\bm{O}^{(k)}_s\right)}{ \sqrt{\det\left(\bm{Q}^{(k)}\right)}},
\eeq 
where as before $\bm{O}^{(k)} = \one - \left( \bm{Q}^{(k)}\right)^{-1}$, $\Tor (\cdot)$ is the \emph{torontonian} \cite{quesada2018gaussian}, and the entries of $S^{(k)}$ take the values $s_i=0$ for no detection and $s_i=1$ for detection. The best known algorithms for computing torontonians have the same complexity as those for calculating hafnians \cite{quesada2018gaussian}.

\subsubsection{Displacements}
The algorithm can also be applied to GBS with displacements, i.e., when the Gaussian state has non-zero mean $\bm{\bar{\alpha}} \neq 0$. We first define the following useful quantities:
\begin{align}
\vec \alpha^{(k)} &= \left(\bm{\bar \alpha}^{(k)},\left[ \bm{\bar \alpha}^{(k)} \right]^*\right), \\
\vec \gamma^{(k)}  &= \left[ \bm{Q}^{(k)} \right]^{-1} \left[\vec \alpha^{(k)} \right]^\dagger, \\
\mathcal{N}^{(k)} &=\frac{\exp\left(-\tfrac{1}{2}  \vec{\alpha}^{(k)}  \left[ \bm{Q}^{(k)} \right]^{-1} \left[ \vec{\alpha}^{(k)} \right]^\dagger \right)}{ \sqrt{\text{det}(\bm{Q}^{(k)})} s_1!\cdots s_k!}.
\end{align}
As shown in Refs.~\cite{bjorklund2018faster,quesada2019franck,quesada2019simulating}, for GBS with displacement the output probabilities are given by
\begin{align}\label{Eq: lhaf}
p(s)  = \mathcal{N}^{(k)} \times  \text{lhaf}\left\{\text{filldiag} \left( {\bm{A}}^{(k)}_s , \vec{\gamma}^{(k)}_s \right)\right\} ,
\end{align}
where $\text{lhaf}(\cdot)$ is the \emph{loop hafnian} introduced in Ref.~\cite{bjorklund2018faster}, $\vec \gamma_s$ is the vector obtained from $\vec \gamma$ by repeating the $i$ and $i+m$  entries of $\vec \gamma^{(k)}$ a total of $s_i$ times, and the function $\text{filldiag} \left( {\bm{A}}^{(k)}_s , \vec{\gamma}^{(k)}_s \right)$ replaces the diagonal of ${\bm{A}}^{(k)}_s$ with the vector $\vec{\gamma}^{(k)}_s$. The loop hafnian has the same complexity as the hafnian, thus its sampling algorithm has the same complexity as GBS without displacement. 
\subsubsection{Linear combinations of Gaussian states}

The algorithm can also be applied to sampling using threshold detectors for states that can be written as linear combinations of Gaussian states:
\begin{align}
\varrho = \sum_{i=1}^\ell q_i \ \varrho_i(\bar{\bm{\alpha}}_i, \bm{\Sigma}_i),
\end{align}
where $\varrho_i(\bar{\bm{\alpha}}_i, \bm{\Sigma}_i)$ is a Gaussian state with 
vector of means $\bar{\bm{\alpha}}_i$ and covariance matrix $\bm{\Sigma}_i$.

The coefficients $q_i$ can form a probability distribution, in which case $\varrho$ is a mixture of Gaussian states, or they can be (possibly negative) real numbers. This last case occurs for instance when describing non-Gaussian states prepared using heralding schemes \cite{quesada2018gaussian}.
If the $q_i$ form a probability distribution, then one only needs to draw the Gaussian state $\varrho_i(\bar{\bm{\alpha}}_i, \bm{\Sigma}_i)$ with probability $q_i$ and directly use the algorithms derived before.  If on the other hand some of the $q_i$ are negative, note that the probability of any event is simply the weighted sum of the probabilities of the Gaussian states $\varrho_i(\bar{\bm{\alpha}}_i, \bm{\Sigma}_i)$; thus, the machinery of the algorithm carries over with minimal modifications. In particular, the time required to generate a sample simply gets multiplied by $\ell$, the number of Gaussian states combined to generate the state $\varrho$.

\subsubsection{Non-negative kernel matrices}

Besides being generalizable to different states and measurements, the algorithm can have polynomial time complexity whenever output probabilities can be estimated efficiently. An important example is when the covariance matrix $\bm{Q}$, or equivalently $\bm{\Sigma}$, is non-negative, i.e., when $\bm{Q}_{ij} \geq 0$ for all $i,j$. Indeed, as we show in Appendix \ref{app:proof}, if $\bm{Q}^{(k)}$ is both non-negative and a proper quantum covariance matrix, then the reduced Kernel matrices $\bm{A}^{(k)}_s$ are also non-negative. This implies that the probabilities $p(S^{(k)})$ can be approximated efficiently since the hafnian of non-negative matrices can be estimated in polynomial time \cite{barvinok1999polynomial, rudelson2016hafnians}.

Let $\bm{G}$ be a skew-symmetric random matrix whose entries above the main diagonal are drawn from the standard normal distribution $\mathcal{N}(0,1)$. Defining a matrix $\bm{W}$ with entries $W_{ij}= G_{ij}\sqrt{a_{ij}}$, it holds that \cite{barvinok1999polynomial, rudelson2016hafnians}
\begin{equation}
\text{Haf}(\bm{A})=\mathbb{E}[\det(\bm W)],
\end{equation}
where $a_{ij}$ are the matrix elements of $\bm{A}$.
To approximate the hafnian of $\bm{A}$, $M$ random matrices $\bm{W}_1, \bm{W}_2, \ldots, \bm{W}_M$ are drawn and used to compute the estimator
\begin{equation}\label{Eq: Haf_approx}
\text{Haf}(\bm{A})\approx \frac{1}{M}\sum_{i=1}^M\det(\bm{W}_i).
\end{equation}
This scenario includes the important situation where the kernel matrix $\bm{A}$ is the adjacency matrix of an unweighted graph.\\

Computing the determinant of a $2n\times 2n$ matrix using the lower-upper (LU) decomposition method takes time $O(n^3)$, and $M$ such determinants have to be calculated to estimate a single hafnian. 
As shown in Appendix \ref{app:proof}, for the cases when the kernel matrix is non-negative and block diagonal $\bm{A} = \bm{B} \oplus\bm{B}$, it follows that the covariance matrix $\bm{Q}$ is also non-negative and thus all probabilities in the algorithm can be estimated in polynomial time up to subexponential errors \cite{rudelson2016hafnians}. From the analysis of the previous section, this leads to an asymptotic running time scaling as $
O\left(m M N^3\right)$.

\section{Benchmarking}
In this section, we test the performance of the algorithm. Tests are based on a Python/C++ implementation, publicly available in the \texttt{samples} module of Xanadu's The Walrus library \cite{code}. C++ code is used to compute hafnians, while all other operations are performed in Python. As stated in section \ref{Sec: complexity}, the runtime of the algorithm scales exponentially with the number of photons and linearly with the number of modes. Since the number of photons is the dominant parameter, we fix the number of modes and test the computational resources required to produce samples with different photon numbers. We consider two scenarios. First, we simulate GBS with equal squeezing levels and an interferometer selected randomly from the Haar measure. Second, we benchmark the approximate sampling algorithm for non-negative kernel matrices built from random Erd\H{o}s-Renyi graphs.

\subsection{Exact algorithm}
When all modes in a GBS device are equally squeezed, the kernel matrix takes the form $\bm{A}=\bm{B}\oplus\bm{B}^*$, with $\bm{B}$ a symmetric matrix. The GBS distribution is then
\beq
p(S) = \frac{1}{\sqrt{\text{det}(\bm{Q})} } \frac{ |\text{Haf}(\bm{B}_S)|^2}{s_1!\ldots s_m!}.
\eeq
For each choice of squeezing parameters and linear interferometer, there exists a corresponding matrix $\bm{B}$. Similarly, for every $\bm{B}$ there is a corresponding setting of interferometer and squeezing parameters \cite{bromley2019applications}. 

\begin{figure}[t!]
	\centering
	\includegraphics[width=0.9\columnwidth]{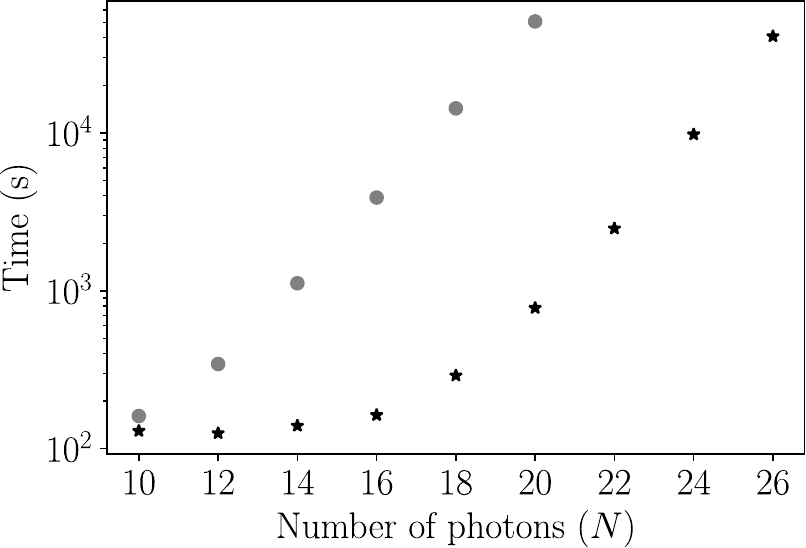}
	\caption{Running times of the exact simulation algorithm for a GBS system with 100 modes. The interferometer is Haar-random and the squeezing levels are set uniformly in the interval $[0,1]$ for each mode. The algorithm is implemented in a cluster of 56  Intel(R) Xeon(R) CPUs operating at 2.6GHz. The top points (circles) indicate the upper bound on the runtime, which occurs when $N$ photons are each observed in the first $N$ modes. The bottom points (stars) show a lower bound on the runtime, when $N$ photons are each observed in the last $N$ modes. Since we fix the detection pattern and also the total number of photons detected, the running time of these extreme cases is independent of the squeezing parameters used for the simulation.}
		\label{Fig:timing_exact}
\end{figure}

We simulate this setting for a GBS device with $m=100$ modes and an interferometer unitary chosen at random from the Haar measure. Squeezing levels in each mode are selected uniformly at random in the interval $[0,1]$. Simulations are performed on a cluster of 56 CPUs. The runtimes for different photon numbers are shown in Fig.~\ref{Fig:timing_exact}. As explained in Sec.~\ref{Sec: complexity}, there are two extreme cases. The fastest runtimes occur when $N$ photons are observed in each of the last $N$ modes. The longest runtimes happen when $N$ photons are each observed in the first $N$ modes. In the simulation, these two situations are hard-coded, giving rise to upper and lower bounds on the total runtime. The lower bound indicates that hours are needed to produce a single sample with 24 or more photons. Proportionate reductions in runtime can be obtained using more powerful processors and parallelizing the computation.

\subsection{Approximate algorithm for non-negative matrices}
\begin{figure}[t!]
	\centering
	\includegraphics[width=0.9\columnwidth]{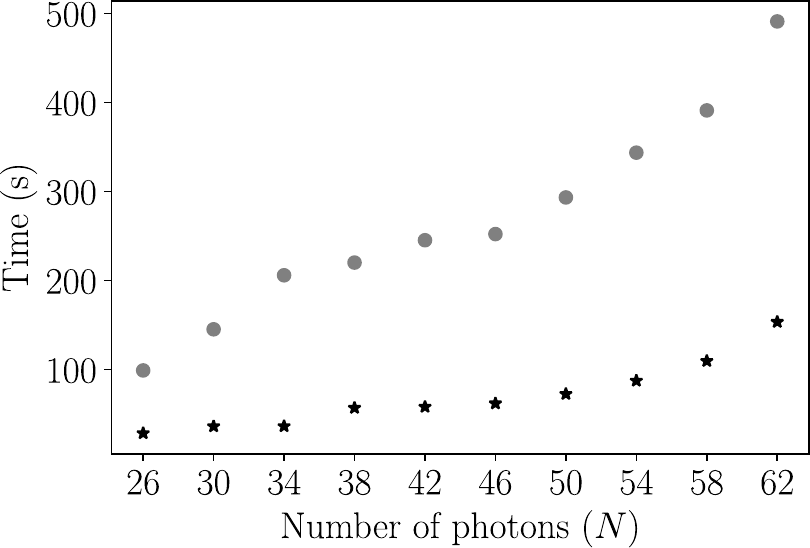}
	\caption{Running times of the approximate simulation algorithm for non-negative kernel matrices. The kernel is the adjacency matrix of a random Erd\H{o}s-Renyi graph with 100 vertices and edge probability $1/2$. The algorithm is implemented in a cluster of 56  Intel(R) Xeon(R) CPUs operating at 2.6GHz. The top points (circles) show the upper bound on the runtime, which occurs when $N$ photons are each detected in the first $N$ modes. The bottom points (stars) indicate a lower bound on the runtime, when $N$ photons are each observed in the last $N$ modes.  }\label{Fig:timing_approx}
\end{figure}

We benchmark the approximate sampling algorithm when the kernels is the adjacency matrix of a random Erd\H{o}s-Renyi graph with 100 vertices, where an edge is added with probability $1/2$. When encoded in a GBS device, this corresponds to $m=100$ modes. In estimating the hafnians using Eq.~\eqref{Eq: Haf_approx}, $M=1000$ samples were used. The mean photon number was set to 20. Upper and lower bounds on the runtimes are shown in Fig.~\ref{Fig:timing_approx}. As expected, the runtimes are considerably shorter for the approximate algorithm, which can for example produce samples with 50 photons in a few minutes. However, this comes at a price: the errors in the approximation are significant.

To monitor the error in the hafnian estimation, we compute the sum of the conditional photon number probabilities when sampling the last mode, i.e., we calculate $\eta=\sum_{i=0}^{n_{\text{max}}} p(s_m=i|s_1,\ldots,s_{m-1})$. For $n_{\text{max}}$ sufficiently large, if all probabilities are calculated correctly, it must hold that $\eta=1$. Therefore the error $\varepsilon=|1-\eta|$ can be used as a simple method to estimate the reliability of the sampling procedure. We observe an average error of $\varepsilon=63\%$ across all samples, with a maximum error of $\varepsilon=228\%$. This can be interpreted as evidence that the algorithm is sampling from a distribution that is far from the target GBS distribution. From standard results in Monte Carlo estimation, the error in the hafnian approximation formula of Eq.~\eqref{Eq: Haf_approx} scales as $1/\sqrt{M}$, meaning that it becomes expensive to decrease the error further. 

\section{Discussion}

We have described an exact classical algorithm for simulating GBS. The algorithm has minimal space complexity and its runtime is directly proportional to the time required to compute output probabilities. These properties make it the best currently known method for simulating GBS devices. Compared to the state of the art in algorithms for simulating Boson Sampling, our results place GBS on similar footing, since in all cases the complexity of the sampling algorithms is proportional to the complexity of computing probabilities. The only difference is that for GBS, the complexity also scales linearly with the number of modes. In our algorithm, this occurs because each mode must be sampled individually to account for the fact that photon numbers can vary among samples, a feature that is not present in Boson Sampling. It remains an open question whether this small dependency on mode number is fundamental or can be removed.

We have implemented the algorithm and benchmarked its runtime as a function of the number of photons. It is informative to make a comparison with the sampling rates of physical devices. Sampling rates in GBS are determined by detector dead times, which limit the speed at which photons can be detected. Depending on specific technologies, detector dead times lead to sampling rates in the range of $10^5$~Hz -- $10^7$~Hz~\cite{hadfield2009single}. On the other hand, even for the approximate algorithm using only $M=1000$ samples for hafnian estimation, sampling rates are on the order of $10^{-2}$ Hz for outputs with a few dozen photons. Even if these rates were boosted by using supercomputing power, the sampling rate of a physical device is likely to be orders-of-magnitude larger than what can be obtained with classical simulators.

Finally, beyond its fundamental importance in determining the practical complexity of GBS simulation, the algorithm can be a valuable tool for researchers working with GBS. The algorithm can be used as a method to test and benchmark new applications, to verify physical implementations, and to model the role of experimental imperfections.

\section*{Acknowledgements}
We thank L. Banchi, T.R. Bromley, B. Gupt, J. Izaac, N. Killoran, A. Mari, H. Qi and M. Schuld for fruitful discussions.

\appendix
\section{Non-negative matrices and GBS}\label{app:proof}
Assume $\bm{\Sigma}$ is a valid quantum complex-normal covariance matrix. This implies that \cite{simon1994quantum}
\eq{
\bm{\Sigma} + \tfrac{1}{2}\bm{Z} \geq 0	\Longrightarrow \bm{\Sigma} > 0,
}
where $\bm{Z} = \left[\begin{smallmatrix}
\one & 0 \\ 0 &-\one
\end{smallmatrix}\right]
$. We furthermore assume that $\bm{\Sigma}$ is non-negative
\eq{
\Sigma_{ij} \geq 0\hspace{5mm} \forall i,j.
}
This implies that $\bm{Q} = \bm{\Sigma}+\one/2$ is also non-negative. We want to show that the kernel matrix $\bm{A} = \bm{X}\left(\one - \bm{Q}^{-1} \right)$ is non-negative. Note that $\bm{X}$ is a permutation matrix and thus $\bm{A}$ is non-negative if and only if 
\eq{\label{QSdef}
\bm{Q}^{-1} = \one - \bm{O},
}
is non-negative.
Since $\bm{Q}$ has non-negative entries and is positive definite, its inverse $\bm{Q}^{-1}$ is a so-called $M$-matrix \cite{poole1974survey}. If $\bm{M}$ is an $M$-matrix, it can be shown that
\eq{
	M_{ii}>0 \text{ and }  M_{ij} \leq 0 \text{ for } i\neq j.
}
Furthermore, if $\bm{M}$ is an $M$-matrix, then there exists a scalar $\lambda_0$  and a matrix $\bm{N}$ such that \cite{poole1974survey}
\eq{\label{Mconst}
	\bm{M} = \lambda_0 \one - \bm{N} , 
}
where $\lambda_0 \geq \max \{\text{eigvals}( \bm{N})\}$ and $N_{ij} \geq 0$. Therefore, since $\bm{Q}^{-1}$ is an $M$-matrix, we can write
\eq{\label{QMmatrix}
\bm{Q}^{-1} = \lambda_0 \one - \bm{N}.
}

Comparing Eq.~\eqref{QSdef} and Eq.~\eqref{QMmatrix}, we can identify 
\eq{\label{ids}
	\lambda_0 & =1 \\
	\bm{N} &= \bm{O}.     
}
However, for this identification to be correct, it remains to show that $\lambda_0=1\geq \max \{\text{eigvals}( \bm{O})\}$.
This is easily seen from the following chain of inequalities
\eq{
	0 &< \bm{\Sigma} 	\Longrightarrow \one/2 < \bm{Q} \Longrightarrow \nonumber\\
	0 &<  \bm{Q} ^{-1} < 2 \one
	\Longrightarrow -\one < \underbrace{ \one - \bm{Q} ^{-1}}_{=\bm{O}} <  \one.
}
Therefore, we conclude that $\bm{O}$ has non-negative entries.
Note that $\bm{Q}$ can also stand for the covariance matrix of a subset of the modes, which in the main text is labelled as $\bm{Q}^{(k)}$. 
This concludes the proof that the kernel matrix is non-negative if the covariance matrix is non-negative.

Now we consider the inverse problem. We study certain kernel matrices that always have non-negative covariance matrices. Consider pure Gaussian states for which the covariance matrix has the following simple parametrization \cite{hamilton2017gaussian,kruse2018detailed}
\eq{
	\bm{Q} &= (\one - \bm{X} \bm{A})^{-1} \\
	\bm{A} &= \left[   \begin{smallmatrix}
		\bm{B} &  0 \\
		0 & \bm{B}^*  
	\end{smallmatrix} \right]\\
	\bm{X} & = \left[   \begin{smallmatrix}
		0 &  \one \\
		\one & 0  
	\end{smallmatrix} \right],
}
where $\bm{B}=\bm{B}^T$ has singular values that satisfy 
\eq{\label{eq:Bbound1}
0 \leq \text{singvals}(\bm{B}) < 1.
} 
As before, let us now specialize to the case where $\bm{B}$ is real and non-negative. The last bound can be restated in terms of the eigenvalues of $\bm{B}$ as
\eq{\label{eq:Bbound}
	-1 < \text{eigvals}(\bm{B}) < 1.
	}

This last fact implies that the spectrum of $\bm{A}$ is also in $(-1,1)$. Using this observation we can rewrite
\eq{
	\bm{Q} &= f(\bm{X} \bm{A}), \text{ where}\\
	f(y) &= \frac{1}{1-y} = \sum_{n=0}^{\infty} y^n.
}
In the last equation we used the bound on the eigenvalues of $\bm{X} \bm{A}$ to expand $f(y)$ as a power series in $y$. The bound in Eq.~\eqref{eq:Bbound} guarantees that the series converges. Now note that $\bm{X}\bm{A}$ has non-negative entries, therefore any of its powers will also have non-negative entries. Furthermore, the sum of two matrices with non-negative entries is also non-negative, thus showing that $\bm{Q}$ is non-negative.

\begin{figure}[t!]
	\includegraphics[width=0.9\columnwidth]{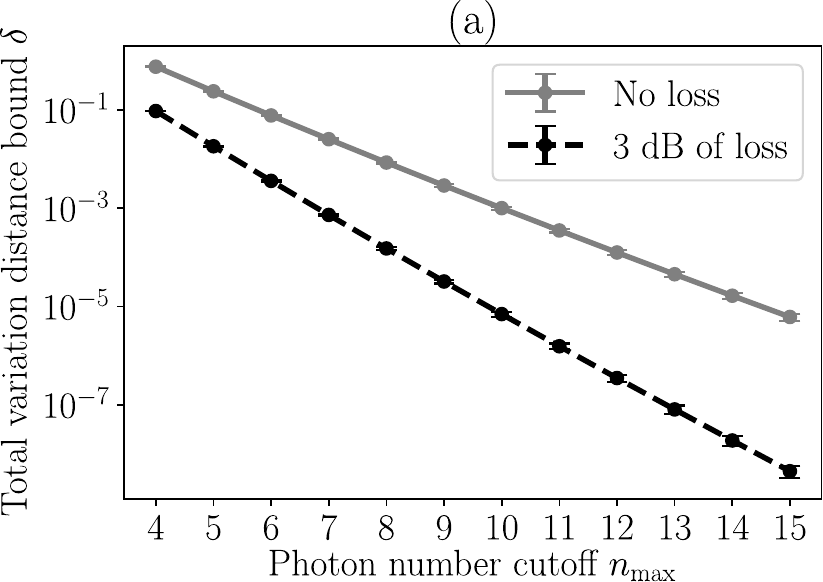}
	\bigskip \\
	\includegraphics[width=0.9\columnwidth]{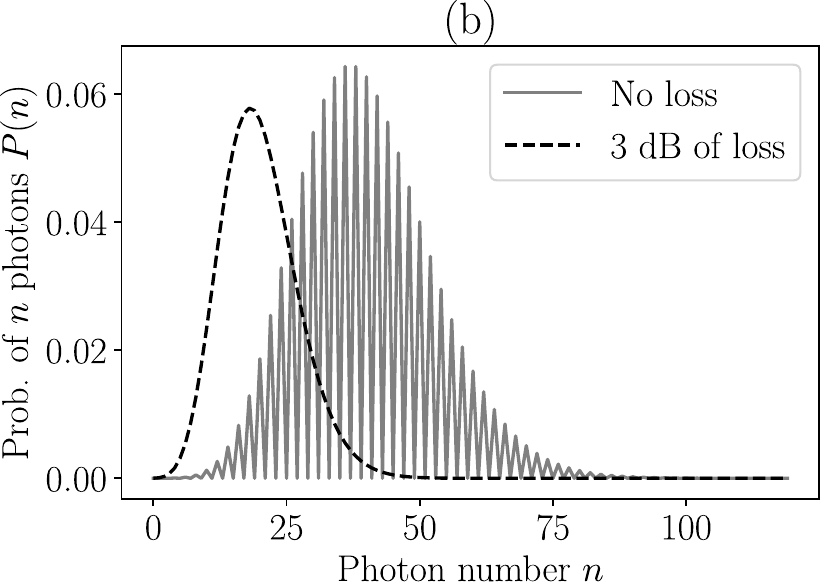}
	\caption{(a) Bound on the total variation distance for a GBS setup as a function of the maximum number of photons $k$ that can be resolved in a detector. We assume 40 squeezed states with input mean photon number $\bar{n}=1$ are sent into a $100 \times 100$ interferometer. 
		The black dotted lines consider the lossless case and the grey lines consider applying 3 dB of loss in each of the modes. Note the exponential decay in the bound on the total variation distance. The results are averaged over 1000 unitaries drawn from the Haar measure and the bars denote one standard deviation.
		(b) We show the distribution of the total photon number $P(n)$ for pure and lossy GBS with the same parameters used in the top figure.
	}\label{Fig:bound}
\end{figure}

\section{Bounding the effect of finite photon-number resolution}\label{app:bound}
Let $S$ denote the random variable describing the outcomes of an ideal GBS setup where the photon-number detectors can resolve any number of photons with perfect accuracy. 
The support of this random variable is $\mathbb{N}_0^m$, where $\mathbb{N}_0$ is the set of natural numbers including zero and $m$ is the number of modes.
Consider the probability distribution $q(S)$ obtained from the ideal distribution $p(S)$ by taking samples $S$ in which more than $n_{\text{max}}$ photons are detected in any of the modes, and mapping them to the event $\varnothing$. For example, for $m=4$ and $n_{\text{max}} = 8$, the sample $(0,1,2,10)$ is mapped to $\varnothing$, whereas the event $(0,1,2,5)$ is mapped to itself. 

Any sample in which no more than $n_{\text{max}}$ photons are detected in each mode has the same probability under the two distributions. Thus, they only differ in events in which more than $n_{\text{max}}$ photons are measured. This allows us to write their total variation distance as follows:
\eq{
\delta &:= \tfrac{1}{2} \sum_{S \in \mathbb{N}_0^m \cup \varnothing} |p(S) - q(S)| = \tfrac{1}{2} \left(q(\varnothing) + \sum_{S > n_{\text{max}}} p(S)\right) \nonumber \\
&=\sum_{S > n_{\text{max}}} p(S). \label{tvd}
}
In the last equation we used the notation $S > n_{\text{max}}$ to indicate any $S$ for which at least one mode had more than $n_{\text{max}}$ photons. The support of the distributions is extended so that $p(\varnothing)=0$ and $q(S)=0$ if $S > n_{\text{max}}$. Finally, we used the fact that $q(\varnothing) = \sum_{S > n_{\text{max}}} p(S)$. 

It then holds that
\begin{align}
\delta =& \sum_{S > n_{\text{max}}} p(S)\nonumber\\
& < \sum_{k=1}^m \sum_{n=n_{\text{max}}+1}^{\infty} P_k(n) = \sum_{k=1}^m \sum_{n=0}^{n_{\text{max}}} (1-P_k(n))\label{bound},
\end{align}
where $P_k(n)$ is the \emph{marginal} probability distribution of the number of photons $n$ in mode $k$.
The quantity in the right hand side of Eq.~\eqref{bound} is a bound and not an equality because events like $(0,1,10,10)$ are counted twice; once in the marginal for the third mode and once in the marginal for the fourth mode (again assuming $m=4$ and $k = 8$).

The single-mode marginal probabilities $P_k(n)$ in Eq. \eqref{bound} can be calculated in cubic time in $n$. This functionality is implemented in the \texttt{quantum} module of The Walrus~\cite{code}), and therefore the bound on the total variation distance can be computed in polynomial time for any $n_{\text{max}}$. In Fig.~\ref{Fig:bound}(a) we plot the values of the bound derived for a GBS setup where 40 squeezed states with input mean photon number $\bar{n}=1$ are sent into a $100 \times 100$ interferometer. For average instances of the interferometer matrix, the bound decays exponentially with the cutoff $n_{\text{max}}$.

Finally, in the case where single-mode Gaussian states are sent into an interferometer one can also calculate the distribution $Q(n)$ of the total photon number by simply taking the convolution of the single-mode photon-number distributions before the interferometer $Q_k(n)$. The distribution of the total photon number is not changed by applying an interferometer, thus $Q(n) = P(n)$ where $P(n)$ is the total photon number distribution after the interferometer. The $k-1$ convolutions of the single mode distributions each require $O\left(n_{\max} \log(n_{\max})\right)$ steps using fast Fourier transforms. This observation allows us to obtain in polynomial time the total photon number distributions shown in Fig.~\ref{Fig:bound}(b) for 40 pure (grey-full lines) or 3 dB-lossy (black-dashed lines) squeezed states sent into a $100 \times 100$ interferometer. Note how loss reduces the mean and variance of the total photon number distribution.
\label{Sec:Appendix2}
\bibliography{gbs}

\end{document}